\documentstyle[prl,aps,epsfig,preprint]{revtex}

\begin{document}

 \draft
\title {{\bf High energy spin excitations in $\bf YBa_2 Cu_3 O_{6.5}$}}

\author{P. Bourges$^{(1)}$, H.F. Fong$^{(2)}$, L.P.
Regnault$^{(3)}$, J. Bossy$^{(4)}$, C. Vettier$^{(5)}$,
D.L. Milius$^{(6)}$,\\ I.A. Aksay$^{(6)}$, and B. Keimer$^{(2)}$
}

\address{1 - Laboratoire L\'eon Brillouin, CEA-CNRS, CE Saclay, 91191 Gif sur Yvette, France}
\address{2 - Department of Physics, Princeton University, Princeton, NJ 08544
USA}
\address{3 - CEA Grenoble, D\'epartement de Recherche Fondamentale sur 
la mati\`ere Condens\'ee, 38054 Grenoble cedex 9, France}
\address{4 - Institut Laue-Langevin, 156X, 
38042 Grenoble Cedex 9, France}
\address{5 - European Synchrotron Research Facility, BP 220, 38043 Grenoble 
cedex, France}
\address{6 - Department of Chemical Engineering, Princeton University, 
Princeton, NJ 08544 USA}

\date{To appear in  Phys. Rev. B (RC), 56, (1997), 1 Nov. }

% \twocolumn[\hsize\textwidth\columnwidth\hsize\csname @twocolumnfalse\endcsname

% 74.20.Mn Nonconventional mechanisms (spin fluctuations, polarons, and
% bipolarons, resonating valence bond model, anyon mechanism, 
% marginal Fermi liquid, Luttinger liquid, etc.)
% 25.40.Fq Inelastic neutron scattering
% 74.72.Bk Y-based cuprates

% Adresse complete LPR :  CEA Grenoble, D\'epartement de Recherche 
% Fondamentale sur la mati\`ere Condens\'ee,  Service de Physique Statistique, 
% Magn\'etisme et Supraconductivit\'e (SPSMS/MDN) 38054 Grenoble cedex 9,France

\maketitle

\begin{abstract}
Inelastic neutron scattering has been used to obtain a
comprehensive description of the absolute dynamical spin susceptibility 
$\chi'' ({\bf q}, \omega)$ of the underdoped superconducting cuprate 
$\rm YBa_2 Cu_3 O_{6.5}$ ($\rm T_c = 52$K) over a wide range of
energies and temperatures (2 meV $\leq$ $\hbar 
\omega$ $\leq$ 120 meV and 5K $\leq$ $T$ $\leq$ 200K). Spin excitations 
of two distinct symmetries (even and odd under exchange of two adjacent 
$\rm CuO_2$ layers) are observed which exhibit {\it two different} 
gap-like features (rather than a single ``spin pseudogap''). The excitations show
dispersive behavior at high energies.
\end{abstract}


%  ]

% \clearpage
The magnetic excitation spectra of doped cuprates contain incisive and highly 
specific information about theories for high temperature superconductivity.
Many models based on strong Coulomb correlations between charge carriers, for 
instance, predict that at high energies the spin dynamics should resemble 
those of the parent antiferromagnetic insulator. Until recently, 
neutron experiments on
metallic and superconducting cuprates were confined to excitation energies 
below $\sim 50$ meV, larger than the superconducting energy gap but
smaller than the intralayer nearest-neighbor superexchange  
$J_{||} \sim 100$ meV which sets the energy scale for spin 
excitations in the undoped antiferromagnetic precursor compounds.
Recent pioneering studies of the single-layer superconductor 
$\rm La_{1.85} Sr_{0.15} CuO_4$ extended these measurements to higher 
energies and established the presence of significant spectral weight at
energies comparable to $J_{||}$ \cite{yamada,haydenLSCO}.
Analogous data for metallic and superconducting $\rm YBa_2 Cu_3 O_{6+x}$ 
(YBCO) have thus far not been reported.

 As YBCO is a bilayer system, such experiments can also answer questions 
about the nature and strength of the coupling between two directly 
adjacent $\rm CuO_2$ layers, a problem of intense current research.
As there are two Cu atoms per unit cell of YBCO, one 
generally expects the formation of bonding and antibonding electronic 
states. Conventional 
band theory predicts the formation of two Fermi surfaces in bands composed 
of bonding and antibonding states. However, theoretical \cite{anderson} 
as well as experimental \cite{ding} evidence indicates that these predictions 
do not generally hold for low-dimensional systems in the presence of 
strong correlations. Here we report neutron scattering measurements that evidence a new  magnon dispersion-like behavior at energies of the order
of $J_{||}$. Moreover, the spin excitation spectrum is characterized by two 
different gap-like features. These features provide fundamental insights 
into the ``spin pseudo-gap'' phenomenon in metallic underdoped  cuprates 
as well as essential new information about the interlayer interactions. 
 
 Transitions between states of the same type (bonding-to-bonding or
antibonding-to-antibonding) and those of opposite types are characterized 
by even or odd symmetry, respectively, under exchange of two
adjacent $\rm CuO_2$ layers. In the cross section for inelastic magnetic 
neutron scattering, these transitions can be distinguished according to their 
distinct dependences on the wavevector component $L$ perpendicular to
the $\rm CuO_2$ sheets \cite{tranquada1,rossat2}: The odd component of the spin 
susceptibility displays a $\sin ^2 (\pi z_{\rm Cu} L)$ dependence whereas the 
even component has the complementary $L$ dependence, $\cos^2 (\pi z_{\rm Cu} L)$. 
(Here, $z_{\rm Cu}=0.285$ is the reduced distance between nearest-neighbor Cu 
spins within one bilayer, and the wavevector ${\bf Q} = (H, K, L)$
is measured in units of the reciprocal lattice vectors $2 \pi/a 
\sim 2\pi/b  \sim 1.63 {\rm \AA}^{-1}$ and $2\pi/c \sim 0.53 {\rm
\AA}^{-1}$). In previous low energy neutron experiments in the metallic 
regime \cite{tranquada1,rossat2,rossat1,mook,lpr,tony1,bourges6.97,tony2,tony6.5,epl,dai}, 
{\it only} spin excitations of odd symmetry were found over 
the entire doping range. Does this reflect a fundamental selection rule, or 
is there a different (higher) energy scale associated with the first type of 
transition? In the latter case, a comparison between the energies and 
susceptibilitites associated with the two excitation modes
would yield quantitative information on 
the strength of the bilayer coupling, obviously an important parameter in 
models of high temperature superconductivity. This issue also bears directly 
on the interpretation of other experiments that are sensitive to this coupling. 
Our new measurements of high energy spin excitations in
underdoped metallic $\rm YBa_2 Cu_3 O_{6.5}$ have much better
resolution and counting statistics than the initial studies on
$\rm La_{1.85} Sr_{0.15} CuO_4$ \cite{yamada,haydenLSCO}.
Excitations of {\it both} odd {\it and} even symmetries are
observed; the even excitations are characterized by a $\sim 53$ meV
energy gap, the odd excitations show a gap-like feature at $\sim 23$
 meV\cite{gap}. Both excitations exhibit large  temperature 
dependences, and a dispersion-like behavior at high energies.

Our sample was a high quality $\rm YBa_2 Cu_3 O_{6.5}$ single crystal of
volume $\sim$ 2.5 cm$^3$ and superconducting transition temperature
(midpoint) $\rm T_c = 52$K. Its preparation and characterization have been
described elsewhere \cite{tony6.5}; in particular, susceptibility measurements shown in
Fig. 2 of Ref. \cite{tony6.5} indicate a sharp superconducting transition (full width 5K) 
which rules out significant inhomogeneities in oxygen content.
The neutron experiments were performed on the triple axis
spectrometers IN8 (installed on a thermal beam) and IN1 (installed on
the hot neutron source) at the Institut
Laue-Langevin (ILL). The incident beam was monochromated
by the (111) reflection of a flat copper crystal on IN8, and by the
(200) or (220) reflections of a vertically curved copper crystal on IN1. 
On both spectrometers, we used a graphite (002) analyzer with 
fixed vertical and adjustable horizontal curvatures. On IN8,
higher-order contamination was eliminated using a pyrolytic 
graphite (PG) filter on the scattered beam for different fixed final energies,
$\rm E_f$= 14.7, 30.5 and 35 meV. On IN1, we worked with 
a fixed final energy of $\rm E_f$= 62.6 meV, and a nuclear resonance 
Er filter was placed on the scattered beam to suppress
contaminations from higher energy neutrons. In all the measurements 
we scanned the wavevector {\bf Q} while keeping the 
energy transfer constant.
Two different scattering geometries were chosen in which wavevectors of the
forms ${\bf Q}=(H,H,L$) or ${\bf Q}=(3H,H,L$), respectively, were accessible 
\cite{tony2}. The results obtained on both instruments and in both geometries
are in good agreement.

Previous experiments
\cite{tranquada1,rossat2,rossat1,mook,lpr,tony1,bourges6.97,tony2,tony6.5,epl,dai} 
have established that the low energy
magnetic cross section is peaked around the in-plane wavevector 
${\bf q}_{\rm 2D} = ({\pi \over a }, {\pi \over b})$, or $H = K = {1 \over 2}$ 
in reciprocal lattice units (r.l.u.). The cross section for odd spin 
excitations exhibits maxima at $L_{\rm odd} \approx 1.7 + 3.5 n$ 
($n$ = integer), and that for even excitations is maximum for
$L_{\rm even} \approx 3.5 n$. We therefore performed constant-energy scans 
along ${\bf Q} = (H, H, L_{\rm odd})$ and $(H, H, L_{\rm even})$, 
shown in Fig. 1. There are numerous
optical phonon modes in the energy range covered by these data.
However, scans in different diffraction zones as well as checks 
against phonon structure factor calculations \cite{tony2} 
established the magnetic 
origin of the peaks. The strong temperature dependence of the peak intensities 
(see below) is also inconsistent with phonon scattering. This methodology 
has previously been applied to neutron data at lower energies
\cite{tranquada1,rossat2,rossat1,lpr,tony1,bourges6.97,tony2,epl}, and 
the results were found to be consistent with polarized beam experiments 
wherever such checks proved feasible \cite{mook,tony2}. The temperature 
dependence of the uniform
background presumably arises from multiphonon scattering.

The scans were fitted to Gaussian profiles convoluted
with the instrumental resolution function, corrected for the Cu magnetic 
form factor \cite{shamoto}, and converted to the dynamical spin 
susceptibility $\chi''$\cite{lovesey} by adjusting for the thermal 
population factor. Fig. 2 shows $\chi_{\rm odd / even}'' (\omega) = 
\int d {\bf q}_{\rm 2D} \chi''({\bf q}_{\rm 2D}, L_{\rm odd / even},
\omega) / \int d {\bf q}_{\rm 2D}$, the susceptibility averaged over
the two-dimensional Brillouin zone in both odd and even channels. [We 
assume an isotropic q-dependence  of the spin susceptibility around 
$(\pi,\pi)$. This assumption is consistent
with the data along (H,H) and a more limited data set along (3H,H).]
While odd excitations are observed over the entire energy range probed by our 
experiments, an energy gap of $\Delta_{\rm even} \sim 53$ meV exists for 
even excitations. Both $\chi_{\rm odd}'' (\omega)$ at low energies and the 
gap for $\chi_{\rm even}'' (\omega)$ are consistent with previous measurements 
in this doping regime \cite{tranquada1,rossat2,lpr} which established a 
lower bound on the gap.
Further, previous studies had given the neutron cross section in
arbitrary units, but many quantitative models now require an
absolute unit scale for $\chi''$. We have therefore calibrated the
magnetic intensity against the phonon spectrum, both against
acoustic phonons at low energies and against an optical phonon at
$\hbar \omega = 42.5$ meV according to a procedure discussed 
elsewhere \cite{tony2}. Both
normalization procedures are in good agreement. 

The temperature dependences of both $\chi_{\rm odd}''$ and $\chi_{\rm
even}''$ are striking. At 200K, $\chi_{\rm odd}'' (\omega)$ shows a broad
peak around 30 meV which sharpens and shifts to lower energy with
decreasing temperature. The $\sim 50$ meV dip in
$\chi_{\rm odd}'' (\omega)$ at low temperatures had not been observed before 
and may be related to the gap in $\chi_{\rm even}'' (\omega)$. Much of the
temperature evolution of $\chi_{\rm odd}'' (\omega)$ takes place in the
normal state; however, there is also an additional enhancement of the peak
intensity at $\rm T_c$ \cite{tony6.5} which is presumably related to the
magnetic resonance peak in the superconducting state dominating the 
magnetic spectra of more heavily doped  samples (x $>$ 0.9) 
\cite{rossat2,rossat1,mook,lpr,tony1,bourges6.97,tony2,tony6.5,epl,dai}. 
While $\chi_{\rm odd}'' (\omega)$ is little affected by
temperature above $\hbar \omega \sim 60$ meV, at these energies 
$\chi_{\rm even}'' (\omega)$ increases by almost a factor of two between 
200K and 5K. In the same temperature interval $\Delta_{\rm even}$ 
softens from $\sim 59$ meV to $\sim 53$ meV. This parallels the enhancement and
softening of the peak in $\chi_{\rm odd}'' (\omega)$ at lower energies.

We now turn to the wavevector dependence of the 
spin susceptibility. At low energies (below $\sim$30 meV), 
the scans are peaked around ${\bf q}_{\rm 2D}$ with an intrinsic width of 
$\Delta {\bf q}_{\rm 2D} \sim 0.2$ \AA$^{-1}$ (FWHM). (Previous
measurements with better {\bf q}-resolution had indicated a flat-topped shape 
of these profiles \cite{sternlieb}). Fig. 1 shows that at 
high energies the peak
broadens to $\Delta {\bf q}_{\rm 2D} \sim 0.3$ \AA$^{-1}$ and disperses away 
from ${\bf q}_{\rm 2D} = ({1 \over 2}, {1 \over 2})$, so that a double peak 
structure emerges. For the sake of simplicity,
we have fitted the {\bf q}-scans to the sum of two displaced
peaks only when a single peak did not give a satisfactory fit. 
The positions and intrinsic widths of the
peaks in $\chi_{\rm odd}'' ({\bf q}_{\rm 2D}, \omega)$ and 
$\chi_{\rm even}'' ({\bf q}_{\rm 2D}, \omega)$ resulting from
this procedure are shown in Fig. 3.

It is instructive to compare these data to previous high energy
measurements on the bilayer antiferromagnetic parent compound 
$\rm YBa_2 Cu_3 O_{6.2}$ and on the single-layer, optimally doped
compound $\rm La_{1.86} Sr_{0.14} CuO_4$. Clearly, the basic
structure of the spin excitation spectrum of $\rm YBa_2 Cu_3
O_{6.5}$ bears some resemblance to the spin wave spectrum of the 
$\rm YBa_2 Cu_3 O_{6.2}$ 
\cite{rossat2,shamoto,tranquada2,dmitry,haydenYBCO}, with the 
ungapped odd (gapped even) excitations 
corresponding to acoustic (optical) spin waves, that is, in-phase (antiphase)
precessions of localized spins in adjacent layers. The dynamical
susceptibility of the  acoustic spin wave mode is 
\begin{equation}
\chi_{\rm odd}'' ({\bf Q}, \omega) \; = \; 4S \pi Z_\chi Z_c \mu_B^2\; 
\frac{1+\gamma({\bf q}_{\rm 2D})}{\sqrt{1-\gamma^2({\bf q}_{\rm 2D})}} \; 
\sin^2 (\pi z_{\rm Cu} L) \; \delta[ \hbar \omega - 4SZ_c J_{||}
\sqrt{1-\gamma^2 ({\bf q}_{\rm 2D})}]
\end{equation}

where $\gamma ({\bf q}_{\rm 2D}) = {1 \over 2} \; [\cos(q_x a)
+ \cos (q_y b)]$, and  the quantum corrections for the spin wave
velocity and the spin susceptibility are taken into 
account respectively as  $Z_c = 1.18$ and $Z_\chi = 0.51$\ following 
theoretical predictions \cite{igarashi} in agreement with
experiment \cite{tony2}. An analogous equation holds for even excitations, 
with $\cos^2 (\pi z_{\rm Cu} L)$ instead of the $\sin^2$ factor and a 
dispersion with a gap of 67 meV \cite{dmitry}. The average of (1) over the 
2D Brillouin zone is finite, $4S Z_\chi/J_{||} \mu_B^2$, and approximately
independent of energy in the energy range probed by our experiment. 
For comparison this quantity is shown as the dashed line in Fig. 2 
(as 10 $\mu_B^2/eV$ with $J_{||}$=102 meV after quantum corrections 
\cite{shamoto}). Clearly, the spectral weights of
spin excitations in antiferromagnetic $\rm YBa_2 Cu_3 O_{6.2}$ and 
superconducting $\rm YBa_2 Cu_3 O_{6.5}$ are comparable. Finally,
the spin wave dispersions of $\rm YBa_2 Cu_3 O_{6.2}$ are
superimposed on the data of Fig. 3. Heuristically,
the odd excitation spectrum of $\rm YBa_2 Cu_3 O_{6.5}$
 can be fitted to a dispersive mode with a pseudo-gap of 23
meV \cite{gap} and a dispersion of $\sim$ 420 meV\AA, 
smaller than the antiferromagnetic spin wave velocity of 650 meV\AA
\cite{shamoto,haydenYBCO}. However, this analysis requires a very
large damping parameter comparable to the gap. 
This pseudo-gap is characteristic of the normal 
state; at this doping level, the dynamical susceptibility is only weakly
modified by superconductivity. Other features of
the spin excitations of $\rm YBa_2 Cu_3 O_{6.5}$ such as the
strong temperature dependence, the pronounced peak-dip
structure at low temperatures and the energy-dependent broadening
of the {\bf q}-width are also very different from spin waves in 
$\rm YBa_2 Cu_3 O_{6.2}$. An explanation of these features
presents a challenge to theories of the spin dynamics in the 
cuprates\cite{noteAF}.

The reports on $\rm La_{1.85} Sr_{0.15} CuO_4$ \cite{yamada,haydenLSCO}
did not contain information about the temperature evolution of
$\chi''(\omega)$, and the measurements did not have sufficient resolution 
to establish the presence or absence of a dip feature in the spectrum.
However, the general shape of 2D {\bf q}-integrated 
$\chi''(\omega)$ in deeply underdoped $\rm YBa_2 Cu_3 O_{6+x}$ 
and optimally doped $\rm La_{2-x} Sr_x CuO_4$ are substantially 
similar. Despite the different {\bf q}-dependences of 
$\chi''({\bf q},\omega)$ at low energies (four sharp incommensurate peaks 
in $\rm La_{1.85} Sr_{0.15} CuO_4$, one broad commensurate peak in 
$\rm YBa_2 Cu_3 O_{6.5}$), both spectra are peaked near 25 meV. 
However, the maximum of $\chi''$ per $\rm CuO_2$ layer is almost 
twice as large in absolute units in $\rm YBa_2 Cu_3 O_{6.5}$
(If given per formula unit, as in Fig. 2, the susceptibility of the
bilayer compound exceeds the one of the single-layer compound by about a 
factor of 4). By contrast,
fully oxygenated $\rm YBa_2 Cu_3 O_{7}$ has a much smaller
normal state susceptibility and a sharp resonance peak in the superconducting
state \cite{mook,lpr,tony1,bourges6.97,tony2}.

In summary, we have reported the first observation of the even part 
of the dynamical spin 
susceptibility of the bilayer superconductor $\rm YBa_2
Cu_3 O_{6.5}$, an important part of the experimental description of spin
fluctuations in the cuprates. Both the odd and the even components are 
characterized by strong and unusual temperature dependences and by a dispersive 
behavior at high energies. Our observation of two different gap-like features in the 
dynamical spin susceptibility demonstrates that the spin-gap phenomenon, a hallmark 
of the underdoped cuprates, is more complex than previously thought.
Our measurements of the spin correlations in absolute units over a 
wide range of energies, wavevectors and temperatures provide an excellent basis for 
theories of this phenomenon.

\vspace{.2in}

\noindent {\bf Acknowledgments}\\

Invaluable technical assistance was provided by P. Palleau, D. Puschner 
and B. Roessli. We wish to thank P.W. Anderson, S. Aubry, A.J. Millis, F. Onufrieva, S. Petit, 
Y. Sidis and C.M. Varma for stimulating discussions. The work at Princeton
University was supported by NSF-DMR94-00362.

\vskip 1 cm

 \subsection*{Figure Captions}

\begin{enumerate}
\item 
Constant energy scans of (a) $\chi_{\rm odd}'' ({\bf Q},\omega)$
through ${\bf Q} = ({1 \over 2}, {1 \over 2}, L_{\rm odd})$
and (b) $\chi_{\rm even}'' ({\bf Q},\omega)$
through ${\bf Q} = ({1 \over 2}, {1 \over 2}, L_{\rm even})$,
measured at 5 K (closed circles) and 200 K (open squares). 
Data have been obtained mostly on IN8  with a final energy of $E_f$=35 meV 
except at the energy transfer of $\hbar \omega$ =15 meV where 
$E_f$=14.7 meV. The scan for $\hbar \omega =80$ meV has been obtained 
on IN1 by rocking the sample  around 
Q=(0.5,0.5,5.4) with $E_f$=62.6 meV. Data obtained on IN8 in different
counting times have been rescaled to the the same monitor 
(monitor=2000, $\approx $ 6 mn/point). Lines are the results of fits to a 
single Gaussian on top of a linear background, 
except at 80 meV where the line corresponds to two identical Gaussian lines 
displaced symmetrically on each side of $H=0.5$.

\item 
Odd (a) and even (b) spin susceptibilities at T=5 K, T=60 K (just above $T_C$) 
and T=200 K  in absolute units (see text). 
Measurements using different final neutron energies $E_f$ obtained 
on the two different 
spectrometers have been rescaled to the same units. The error bars do not
include a $\sim 30$\% normalization error.
\item 
Spin excitation spectrum for odd (open symbols) 
and even (closed circles) excitations at 5 K. The open square 
indicates the energy of the maximum of the odd susceptibility. 
The horizontal bars represent the full width at half maximum after a 
Gaussian deconvolution from the spectrometer resolution. The dotted 
lines correspond to the spin-wave dispersion relation in the insulating antiferromagnetic state with $J_{||}=$ 120 meV 
and $J_{\perp}/J_{||}$= 0.08 (without quantum corrections)
\cite{dmitry,haydenYBCO}. 
\end{enumerate}

% \clearpage
% \clearpage

\begin{figure}

\centerline{\epsfig{file=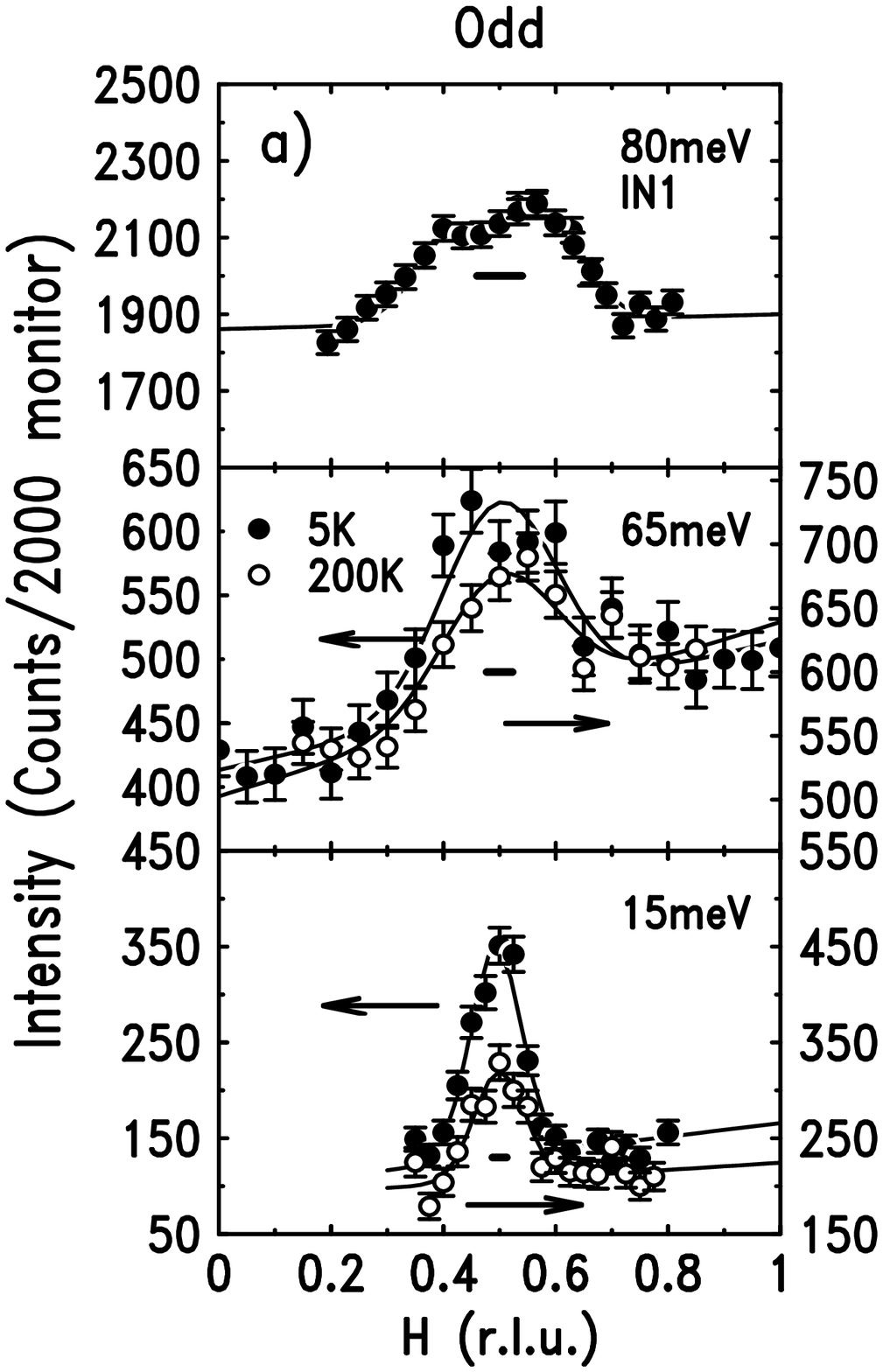,height=15 cm,width= 12 cm}}
\centerline{\epsfig{file=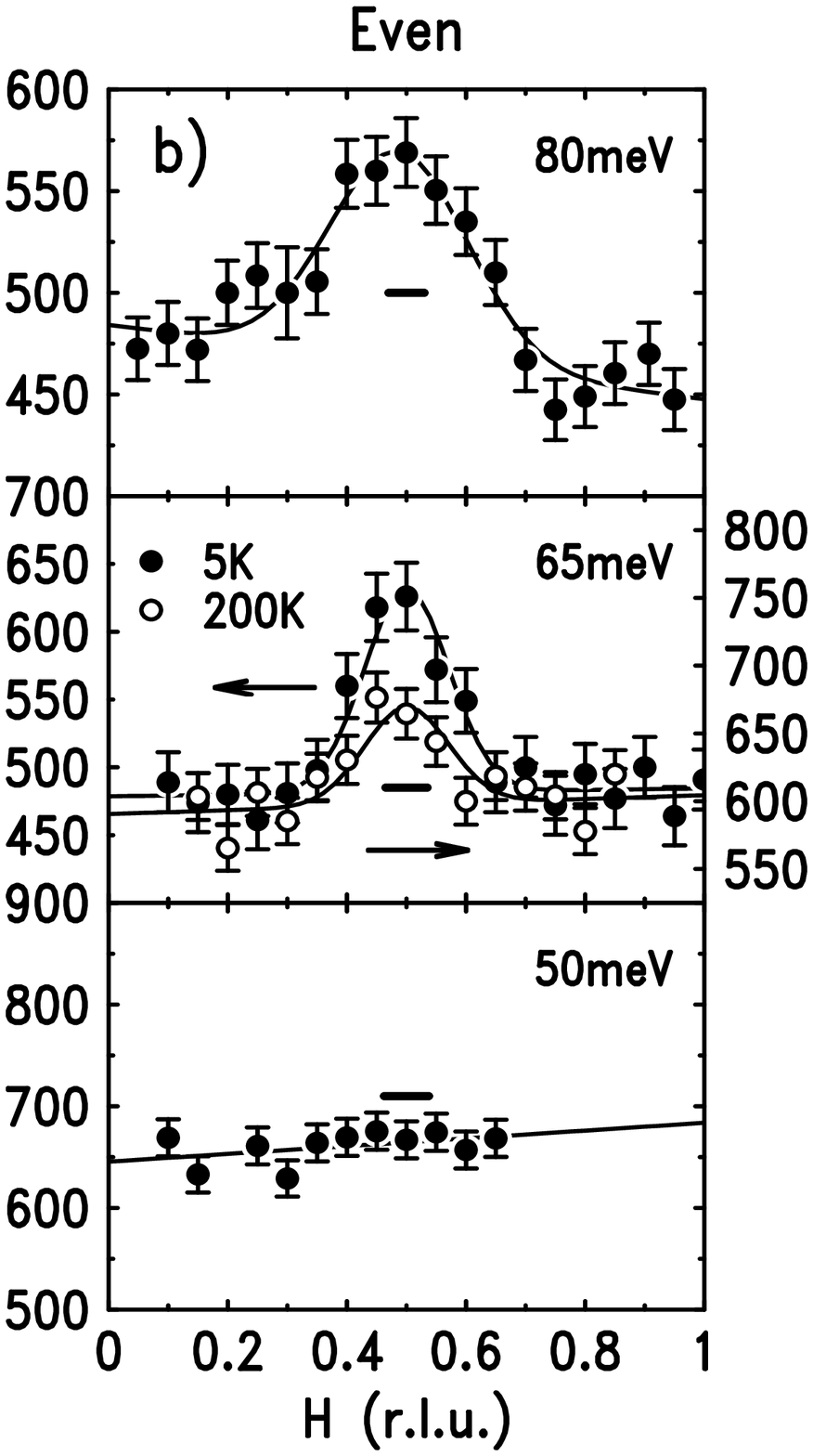,height=15 cm,width= 12 cm}}

\label{fig1} \end{figure}

\clearpage

\begin{figure}

\centerline{\epsfig{file=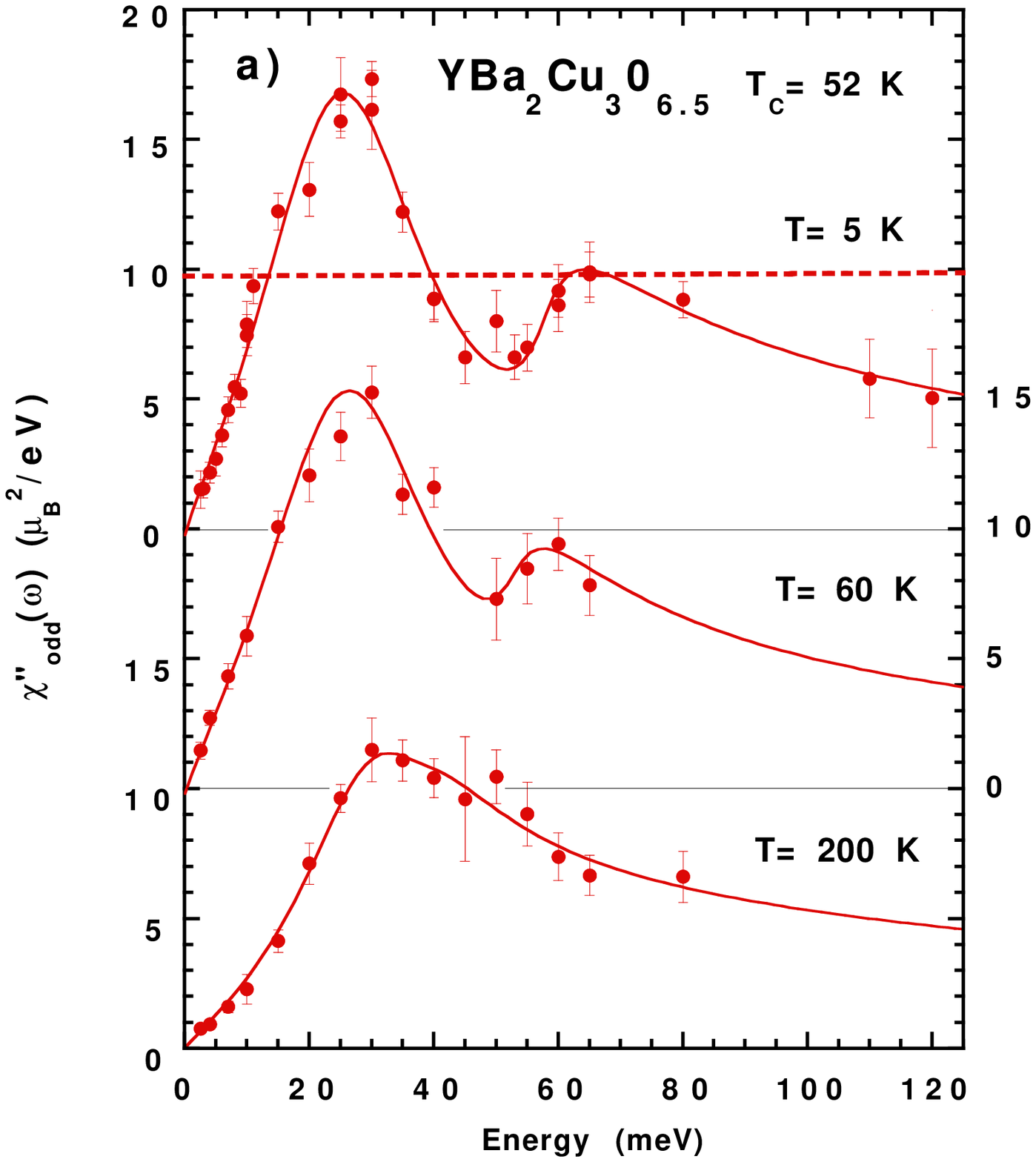,height=16 cm,width= 13 cm}}

\label{fig2a} \end{figure}

\begin{figure}

\centerline{\epsfig{file=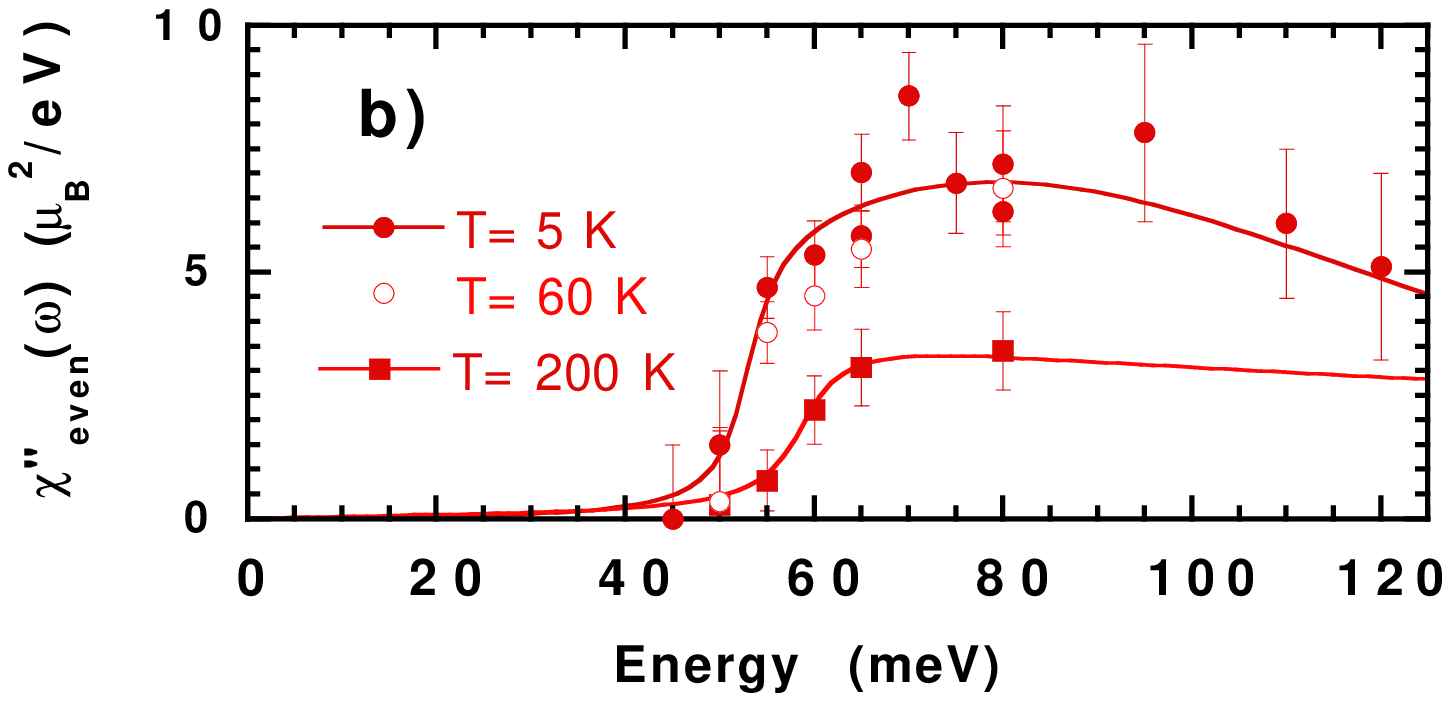,height=6 cm,width= 12 cm}}

\label{fig2b} \end{figure}

\clearpage

\begin{figure}
\centerline{\epsfig{file=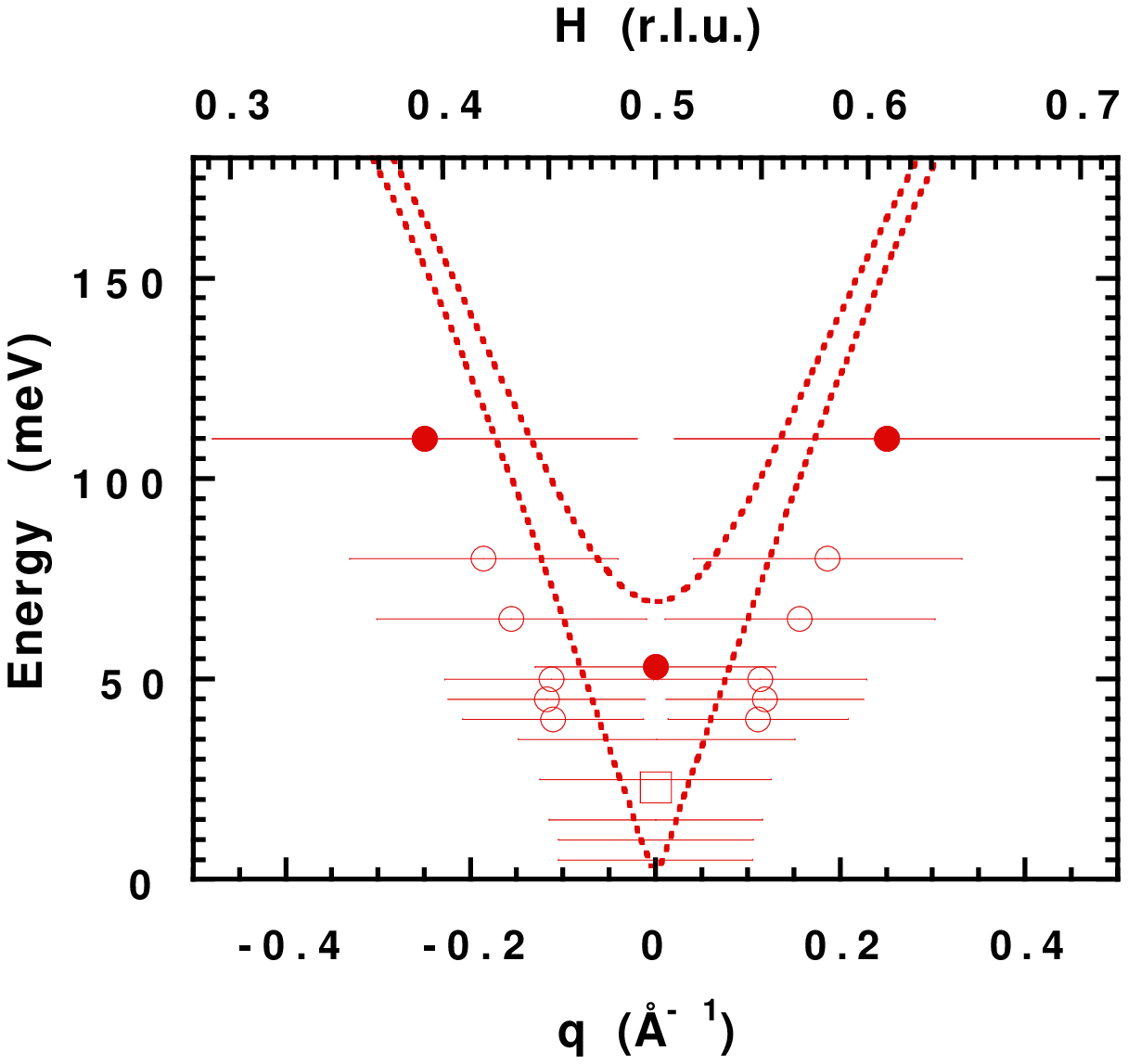,height=12 cm,width= 12 cm}}

\label{fig3} \end{figure}


\begin{thebibliography}{99}
\bibitem{gap} There is no unique way of specifying the pseudo-gap magnitude.
Here, the spin pseudogap energy is identified as the energy at which
dynamical susceptibility is maximum. This definition then differs
from previous conventions \cite{rossat1}. The normal-state spin pseudogap also differs from the true spin gap observed only in the  superconducting state which occurs around 2-5 meV for $x \simeq 0.5$ \cite{rossat2,lpr}.
\end{thebibliography}
\end{document}